\documentclass[14pt]{article}

\usepackage{amsfonts}
\usepackage{epsfig}
\usepackage{a4}
\usepackage{theorem}
\usepackage{amssymb}
\usepackage{graphicx}
\input axodraw.sty

\newcommand{\be}[1]{\begin{equation}\label{#1}}
\newcommand{\ee}{\end{equation}}
\newcommand{\ba}[1]{\begin{eqnarray}\label{#1}}
\newcommand{\ea}{\end{eqnarray}}
\begin{document}
\title{WILSON COEFFICIENT FOR VACUUM CONDENSATE OF PHOTONS FROM
ELECTRODYNAMICS PHOTON PROPAGATOR}
\author{
A.R.~Fazio
\vspace{3mm} \\
\small School of Physics and Centre for Theoretical Physics\\
\small University of Witwatersrand, Johannesburg\\
\small Wits 2050, South Africa\\
\small fazioA@physics.wits.ac.za\vspace{3mm}\\}
\maketitle
\begin{abstract}
In the framework of the operator product expansion I provide a
detailed evaluation of the contribution of the lowest photon
condensate to the two-point photon Green function for
four-dimensional spinor electrodynamics and three-dimensional
scalar electrodynamics. Since the above mentioned condensate
affects the part of the propagator contributing to physical
amplitudes I suggest that this condensate could play a role in the
non-perturbative dynamics of Abelian gauge field theories.
\end{abstract}

\section{Introduction}
Matrix elements of composite operators can bring non-perturbative
information about the dynamics of some quantum field model. It was
pointed out in \cite{Jackiw,Pisarsky} that superrenormalizable
theories cure their infrared divergences if one allows for
coupling constant logarithms in the perturbative expansion with
remaining non-perturbative terms determined by matrix elements of
composite operators. This analysis was performed for three
dimensional electrodynamics, either scalar or spinor, and for both
of the cases it was concluded that the vacuum expectation value of
the square of the photon field, $<A^2>$, provides a
non-perturbative contribution to the self-energy of matter fields.
It was moreover argued that due to the lack of gauge invariance
this expectation value cannot contribute to photon propagator.
Since recently it has been proved the gauge invariance of $<A^2>$
in gauge field theories either Abelian or non-Abelian
\cite{slavnov}, I will check if the above mentioned condensate can
affect physical gauge invariant amplitude in which photon
propagator is sandwiched between conserved currents. In this
letter I will calculate the Wilson coefficient for $<A^2>$ from
the operator product expansion of photon propagator in
four-dimensional scalar electrodynamics and three dimensional
scalar electrodynamics \cite{wilson}. I will use the plane wave
method \cite{vari}, shown to be powerful for QCD-sum rules but
totally equivalent to other methods present on the market, as for
example the Covariant gauge coordinates space method.
\cite{scadron}

\section{Wilson coefficient for $ <A^2>$
in four-dimensional spinor electrodynamics}

I consider massive fermions $\psi$ interacting with a massless
Abelian gauge field $A_\mu$ in four dimensions Minkowskyan
space-time ,
\begin{eqnarray}
S_{\rm QED}&=&\int d^4 x \left[\imath
\bar{\psi}(\widehat{\partial_\mu}-\imath
e \widehat{A_\mu}-m)\psi-\frac{1}{4} F^{\mu\nu}F_{\mu\nu}\right]\\
(\widehat{\partial_\mu}-\imath e
\widehat{A_\mu})&=&\gamma^\mu(\partial_\mu-\imath e
A_\mu)\nonumber\\
F^{\mu\nu}&=&\partial_\mu A_\nu-\partial_\nu A_\mu \nonumber
\end{eqnarray}
The coupling $e$ is dimensionless; the interaction is
renormalizable.

The aim of this section is to compute the coefficient of the
vacuum condensate $<0|A^2|0>$ in the operator product expansion,
as $x\rightarrow0$, of
\begin{equation}
<0|T A_\mu (x) A_\nu (0)|0>\sim \sum_{O} C_O (x^2)<0|[O(0)]|0>.
\end{equation}
where $[O(0)]$ denotes some renormalized operator.

First I consider the tree graphs for
\begin{equation}
<0|T A_\mu (x) A_\nu(0) A_\tau (p_1)A_\sigma (p_2)|0>.
\end{equation}
These give
\begin{equation}
\frac{\imath^2}{p_1^2 p_2^2}(g_{\mu\tau}g_{\nu\sigma}\exp(\imath
p_1 x)+g_{\mu\sigma}g_{\nu\tau}\exp(\imath p_2 x))
\end{equation}
and since I am interested in the insertion of $<A^2>$ in the
photon propagator, I will consider the previous amplitude after
the contraction with $g_{\mu\nu}$
\begin{equation}
\frac{\imath^2}{p_1^2 p_2^2}g_{\sigma\tau}(\exp(\imath p_1
x)+\exp(\imath p_2 x)).
\end{equation}
Expansion in a power series about $x=0$ gives
\begin{equation}
\frac{\imath^2}{p_1^2 p_2^2}g_{\sigma\tau}(2+ \imath
(p_1+p_2)x+.....),
\end{equation}
this is equivalent to replace
\begin{equation}
T A_\mu (x) A_\nu (0)= \frac{g_{\mu\nu}}{4} A^2(0)+
\frac{g_{\mu\nu}}{8} x^\rho\partial_\rho A^2......\label{OPE}
\end{equation}
Since (\ref{OPE}) is an operator equality, it is valid for all its
matrix elements, in particular the vacuum expectation value
$<A^2>$ affects photon propagator.

The first non trivial contribution to the two-point function of $T
A_\mu (x) A_\nu (0)$ is $e^4$-order and it is given by the
following sum of six Feynman diagrams:
\begin{eqnarray}
&&-\frac{e^4}{24p_1^2 p_2^2}\int\frac{d^4 q}{(2\pi)^4}\frac{d^4
k}{(2\pi)^4}\frac{\exp(-\imath q x)}{q^2 (q+p_1+p_2)^2}\nonumber\\
&&\left\{{\rm
Tr}\left[\gamma_\nu\frac{1}{\widehat{k}-\widehat{q}-m}
\gamma_\mu\frac{1}{\widehat{k}-m}
\gamma_{\sigma}\frac{1}{\widehat{k}+\widehat{p_2}-m}
\gamma_\tau\frac{1}{\widehat{k}+\widehat{p_1}+\widehat{p_2}-m}\right]
\right.\nonumber\\ &&\left.+{\rm Tr}\left[\gamma_\nu\frac{1}
{\widehat{k}+\widehat{p_1}+
\widehat{p_2}+\widehat{q}-m}\gamma_\mu\frac{1 }
{\widehat{k}+\widehat{p_1}+\widehat{p_2}-m}
\gamma_{\tau}\frac{1}{\widehat{k}+\widehat{p_2}-m}
\gamma_\sigma\frac{1}{\widehat{k}-m}\right] \right.\nonumber\\
&&\left.+{\rm
Tr}\left[\gamma_\tau\frac{1}{\widehat{k}-\widehat{q}-m}
\gamma_\mu\frac{1} {\widehat{k}-m}
\gamma_{\sigma}\frac{1}{\widehat{k}+\widehat{p_2}-m}
\gamma_\nu\frac{1}{\widehat{k}-\widehat{q}-\widehat{p_1}-m}
\right] \right.\nonumber\\ &&\left.+{\rm
Tr}\left[\gamma_\nu\frac{1}{\widehat{k}-\widehat{p_2}-\widehat{q}-m}
\gamma_\sigma\frac{1} {\widehat{k}-\widehat{q}-m}
\gamma_{\mu}\frac{1}{\widehat{k}-m}\gamma_\tau\frac{1}{\widehat{k}+\widehat{p_1}
-m}\right] \right.\nonumber\\ &&\left.+{\rm
Tr}\left[\gamma_\nu\frac{1}{\widehat{k}-\widehat{p_1}-\widehat{p_2}-\widehat{q}
-m}\gamma_\sigma\frac{1} {\widehat{k}-\widehat{q}-\widehat{p_1}-m}
\gamma_{\tau}\frac{1}{\widehat{k}-\widehat{q}-m}
\gamma_\mu\frac{1}{\widehat{k}-m}\right] \right.\nonumber\\
&&\left.+{\rm
Tr}\left[\gamma_\nu\frac{1}{\widehat{k}-\widehat{p_1}-\widehat{p_2}-\widehat{q}
-m}\gamma_\tau\frac{1} {\widehat{k}-\widehat{q}-\widehat{p_2}-m}
\gamma_{\sigma}\frac{1}{\widehat{k}-\widehat{q}-m}
\gamma_\mu\frac{1}{\widehat{k}-m}\right] \right\}.\label{box}
\end{eqnarray}
The expression used here and in the following for the external
legs of the photon propagator,
\begin{equation}
\frac{\imath g_{\mu\nu}}{q^2},
\end{equation}
doesn't take into account gauge-dependent terms, because they
don't contribute to $S-$matrix elements when photon-propagator is
sandwiched between two conserved currents.

When $x\rightarrow0$ the integral on $q$ in (\ref{box}) diverges
as a logarithm, since, as it will be proved, the integral on $k$
is ultraviolet and infrared finite. This logarithmic divergence is
a symptom of the fact that there are two important region of $q$
that contribute. The first is where $q$ is finite as $x\rightarrow
0$, the graphs receive contribution by replacing $T A_\mu (x)
A_\nu (0)$ by $\frac{g_{\mu\nu}}{4}A^2(0)$. The second region is
where $q$ becomes large, up to $O(1/x)$ as $x\rightarrow 0$; in
this region the interaction vertex in coordinate space is close to
$x$ and 0. In the second region the loop is confined to a small
region in coordinate space. From the point of view of $p_1$ and
$p_2$ the loop is a point, therefore we should be able to
represent the contribution of this region by an extra term in the
Wilson coefficient of $A^2$:
\begin{equation}
T A_\mu(x)A_\nu(0)\sim C_{A^2}(x)\frac{g_{\mu\nu}}{4}[A^2(0)]+....
\end{equation}
\begin{equation}
C_{A^2}(x)= 1+ e^4 c(x^2)
\end{equation}
I shall calculate $c(x^2)$. Now the contribution of the first
region is given by replacing $T A_\mu(x)A_\nu(0)$ by
$\frac{g_{\mu\nu}}{4}[A^2(0)]$. So let me add and subtract the
renormalized Green function of the renormalized operator
$[A^2(0)]$:
\begin{equation}
\frac{g_{\mu\nu}}{4}<0|T[A^2 (0)] A_\tau(p_1)A_\sigma(p_2)|0>.
\end{equation}
The Green function provide the contribution to the Wilson
coefficient of order $1$ from the first region. The reminder
contribution of order $e^4$ is:
\begin{eqnarray}
&&-\frac{e^4}{48p_1^2 p_2^2}\int\frac{d^4q}{(2\pi)^4}\frac{d^4
k}{(2\pi)^4}\frac{\exp(-\imath q x)-1}{(q^2)^2(k-q)^2
(k^2)^3}\times\nonumber\\ && {\rm
Tr}\left[\gamma^\mu(\widehat{k}-\widehat{q})\gamma_\mu
\widehat{k}\gamma_\sigma \widehat{k}\gamma_\tau \widehat{k}+
\gamma^\mu(\widehat{k}-\widehat{q})\gamma_\mu
\widehat{k}\gamma_\tau \widehat{k}\gamma_\sigma
\widehat{k}\right]+
\label{c_1}\\
\nonumber && -\frac{e^4}{96p_1^2
p_2^2}\int\frac{d^4q}{(2\pi)^4}\frac{d^4
k}{(2\pi)^4}\frac{\exp(-\imath q x)-1}{(q^2)^2 [(k-q)^2]^2
(k^2)^2}\times\\ \nonumber && {\rm
Tr}\left[\gamma^\mu(\widehat{k}-\widehat{q})\gamma_\tau
(\widehat{k}-\widehat{q})\gamma_\mu \widehat{k}\gamma_\sigma
\widehat{k}+ \gamma^\mu (\widehat{k}-\widehat{q})\gamma_\sigma
(\widehat{k}-\widehat{q})\gamma_\mu \widehat{k}\gamma_\tau
\widehat{k}\right]\nonumber
\end{eqnarray}
where it has been put $p_1=p_2=m=0$, for the reason explained in
the following.

Since to the lowest order
\begin{equation}
<0|T A^2 A_\tau(p_1)A_\sigma(p_2)|0>=\frac{-2
g_{\sigma\tau}}{p_1^2 p_2^2},
\end{equation}
it is possible to identify $c(x^2)$ as the $x\rightarrow 0$
behaviour of (\ref{c_1}). That leading power behaviour is
independent on $p_1$, $p_2$, $m$, as it can be easily seen by
differentiating (\ref{c_1}) with $p_1$, $p_2$, $m \neq 0$, with
respect to any of these variables. The result is a convergent
integral which goes to zero like a power of $x$ when $x\rightarrow
0$. The definition of $c(x^2)$ is therefore at $p_1=p_2=m=0$.

After making some $\gamma-$gymnastic (\ref{c_1}) becomes:
\begin{eqnarray}
&&-\frac{e^4}{24p_1^2 p_2^2}\int\frac{d^4q}{(2\pi)^4}\frac{d^4
k}{(2\pi)^4}\frac{\exp(-\imath q x)-1}{(q^2)^2 (k-q)^2
(k^2)^3}\times\nonumber\\ && [8 g_{\sigma\tau}k^4 -16 k^2
k_{\sigma}k_{\tau}-8g_{\sigma\tau}k^2(k q)+32 (k q) k_\sigma
k_\tau-8k^2 k_\tau q_\sigma-8k^2 k_\sigma q_\tau]\label{c_2}\\
&&\nonumber -\frac{e^4}{48p_1^2
p_2^2}\int\frac{d^4q}{(2\pi)^4}\frac{d^4
k}{(2\pi)^4}\frac{\exp(-\imath q x)-1}{(q^2)^2 ((k-q)^2)^2
(k^2)^2}\times\\&&[16 k^2 (k-q)_\tau(k-q)\sigma-16 k_\sigma k_\tau
(k-q)^2-8g_{\sigma\tau}k^2 (k-q)^2]\label{c12}.
\end{eqnarray}
Let me fix on $k-$integral. A first contribution to the sum
(\ref{c_2}) and (\ref{c12}) is given by the sum of potential
logarithmic divergent terms:
\begin{equation}
-\frac{e^4}{24p_1^2 p_2^2}\int\frac{d^4 k}{(2\pi)^4}\left[\frac{8
g_{\sigma\tau}k^4-16 k^2 k_\sigma k_\tau}{(k^2)^3
(k-q)^2}-\frac{4g_{\sigma\tau} (k^2)^2}{(k^2)^2
[(k-q)^2]^2}\right].
\end{equation}
Integrals can be performed by introducing the Feynman-parameter
$x$
\begin{eqnarray}
&&-\frac{\imath e^4 q^2}{384\pi^2 p_1^2
p_2^2}\left(-\frac{1}{2}\times(-8)-4\right)
\lim_{\varepsilon\rightarrow 0}\Gamma(\varepsilon)\\
&&-\frac{\imath e^4 q^2}{48\pi^2 p_1^2 p_2^2}\int_{0}^{1}dx
(x-1)\left[\frac{x^2 g_{\sigma\tau}}{(x^2 q^2-x q^2)}-2\frac{x^2
q_\sigma q_\tau}{(x^2 q^2-x q^2)}\right]
\end{eqnarray}
and logarithmic divergences cancel each others \cite{QED}, giving
the finite result:
\begin{equation}
-\frac{\imath}{4\pi^2}\frac{e^4}{p_1^2
p_2^2}\left(g_{\sigma\tau}-2\frac{q_\sigma q_\tau}{q^2}\right).
\end{equation}
The following finite contribution remains in (\ref{c_2}):
\begin{eqnarray}
-\frac{e^4}{24p_1^2 p_2^2}&&\int\frac{d^4 k}{(2\pi)^4}\frac{1}{
(k^2)^3 (k-q)^2}\times\nonumber\\&&\left[-8g_{\sigma\tau}k^2(k
q)+32 k_\sigma k_\tau (k q)-8k^2k_\tau q_\sigma-8k^2k_\sigma
q_\tau\right]
\end{eqnarray}
and turning these integrals into a Gaussian form, after using
Feynman parameters one easily gets:
\begin{equation}
\frac{\imath e^4}{96\pi^2 p_1^2 p_2^2}g_{\sigma\tau}.
\end{equation}
The remaining terms in (\ref{c12}) don't add any contributions
because:
\begin{equation}
\int\frac{d^4 k}{(2\pi)^4}\left[\frac{16k^2(k-q)_\sigma(k-q)_\tau}
{(k^2)^2[(k-q)^2]^2}- \frac{16(k-q)^2 k_\sigma k_\tau}
{(k^2)^2[(k-q)^2]^2}\right]= 0,
\end{equation}
as it can be easily seen by making the change of variables
$k\rightarrow k+q$ and observing that the two integrals are even
functions of the four-vector $q$ or by explicit integration using
Feynman parameters. Moreover
\begin{eqnarray}
&&-\frac{e^4}{48p_1^2p_2^2}\int\frac{d^4k}{(2\pi)^4}
\frac{-8g_{\sigma\tau}k^2 q^2+16g_{\sigma\tau}k^2 (k q)}{(k^2)^2
[(k-q)^2]^2}=\nonumber\\
&&-\frac{g_{\sigma\tau}e^4}{48p_1^2p_2^2}\int\frac{d^4k}{(2\pi)^4}\frac{1}
{k^2(k-q)^2}+\frac{g_{\sigma\tau}e^4}{48p_1^2p_2^2}
\int\frac{d^4k}{(2\pi)^4}\frac{1}{[(k-q)^2]^2}=\nonumber\\
&&\frac{\imath}{16 \pi^2}\lim_{\varepsilon\rightarrow
0}\Gamma(\varepsilon)-\frac{\imath}{16
\pi^2}\lim_{\varepsilon\rightarrow 0}\Gamma(\varepsilon)=0,
\end{eqnarray}
where last line has been obtained after the integration through
Feynman-parameters.
 The sum of all contributions is:
\begin{eqnarray}
&&\frac{2\imath e^4}{192\pi^2 p_1^2 p_2^2}\int\frac{d^4
q}{(2\pi)^4}\frac{\exp(-\imath q x)-1}{(q^2)^2}
\frac{q_\sigma q_\tau}{q^2}\nonumber \\
&&=\frac{2\imath e^4 g_{\sigma\tau}}{16\pi^2 p_1^2
p_2^2}\int\frac{d^4 q}{(2\pi)^4}\frac{\exp(-\imath q
x)-1}{(q^2)^2}+ ......,
\end{eqnarray}
where dots indicate terms vanishing as $x\rightarrow0$.

The integral can be done by using
\begin{equation}
\frac{1}{(q^2)^2}=\int_{0}^{+\infty} dz \exp(-z(-q^2))z
\end{equation}
The $e^4$-order Wilson coefficient for $<A^2>$ in four-dimensional
spinor quantum electrodynamics is:
\begin{equation}
1+\frac{\imath e^4}{6144
\pi^4}\left(\gamma+\log\left(-\frac{x^2}{4}\right)\right).
\end{equation}

\section{Wilson coefficient for $<A^2>$
in three-dimensional scalar electrodynamics}

In this section I will confirm a non-zero result for the Wilson
coefficient of $<A^2>$, also for three-dimensional scalar
electrodynamics. Although the logic that I am going to follow is
the same than in the previous section, here I will work in the
momentum representation. Divergent integral will be regularized by
analytical continuation as explained in the note at the end of
this section.

The action is
\begin{eqnarray}
S=\int d^3x
\left[-\frac{1}{4}F_{\mu\nu}F^{\mu\nu}+|(\partial_{\mu}+\imath e
A_\mu)\varphi|^2\right],
\end{eqnarray}
which describes the dynamics of massless scalar field $\varphi$
interacting with photons $A_\mu$. The square of the coupling
constant $e$ ha dimensions of mass; the interaction is
super-renormalizable.

The Wilson coefficient for $<A^2>$ is again obtained from the
$q^2\rightarrow \infty$ behaviour of the amplitude:
\begin{equation}
g^{\mu\nu}<0|T A_\mu (q) A_\nu(0) A_\tau (p_1)A_\sigma (p_2)|0>.
\label{master}
\end{equation}

To lowest order one has
\begin{equation}
<0|T A^2 A_\tau(p_1)A_\sigma(p_2)|0>=\frac{-2
g_{\sigma\tau}}{p_1^2 p_2^2}.
\end{equation}
The higher corrections to the Green function (\ref{master})
involve three kinds of diagrams.

The sum of diagrams of the first type gives:
\begin{eqnarray}
&&\frac{4 e^4 g_{\tau\sigma}}{p_1^2
p_2^2}\frac{1}{(q^2)(q+p_1+p_2)^2}\int\frac{d^3
k}{(2\pi)^3}\frac{1}{k^2(k-p_1-q)^2}+\nonumber\\
&& \frac{4 e^4 g_{\tau\sigma}}{p_1^2
p_2^2}\frac{1}{(q^2)(q+p_1+p_2)^2}\int\frac{d^3
k}{(2\pi)^3}\frac{1}{k^2(k-p_2-q)^2}+\nonumber\\
&& \frac{2 e^4 g_{\tau\sigma}}{p_1^2
p_2^2}\frac{1}{(q^2)(q+p_1+p_2)^2}\int\frac{d^3
k}{(2\pi)^3}\frac{1}{k^2(k+p_1+p_2)^2}.\label{K1}
\end{eqnarray}
It contributes to Wilson coefficient by
\begin{equation}
-\frac{e^4}{2}\frac{1}{(q^2)^2\sqrt{q^2}}.
\end{equation}
All integrals are convergent.The last term in (\ref{K1})
contributes to the Wilson coefficient of the expectation value of
a non-local operator determined by the action of
$\frac{1}{\sqrt{(p_1+p_2)^2}}$ on $[A^2(0)]$.

The sum of diagrams of second type is:
\begin{eqnarray}
&&-\frac{2 e^4 g_{\sigma\tau}}{p_1^2 p_2^2}\frac{1}{q^2
(q+p_1+p_2)^2}\int\frac{d^3 k}{(2\pi)^3}\frac{(2k-q)_\mu
(2k-q-p_1-p_2)^\mu}{k^2(k-q-p_1-p_2)^2(k-q)^2}\nonumber\\
&&-\frac{2 e^4}{p_1^2 p_2^2}\frac{1}{q^2
(q+p_1+p_2)^2}\int\frac{d^3 k}{(2\pi)^3}\frac{(2k+p_1)_\tau
(2k-q)_\sigma}{k^2(k+p_1)^2(k-q)^2}\nonumber\\
&&-\frac{2 e^4}{p_1^2 p_2^2}\frac{1}{q^2
(q+p_1+p_2)^2}\int\frac{d^3 k}{(2\pi)^3}\frac{(2k+p_2)_\sigma
(2k-q)_\tau}{k^2(k+p_2)^2(k-q)^2}\nonumber\\
&&-\frac{2 e^4}{p_1^2 p_2^2}\frac{1}{q^2
(q+p_1+p_2)^2}\int\frac{d^3 k}{(2\pi)^3}\frac{(2k+p_1)_\tau
(2k+q+p_1+p_2)_\sigma}{k^2(k+p_1)^2(k+q+p_1+p_2)^2}\nonumber\\
&&-\frac{2 e^4}{p_1^2 p_2^2}\frac{1}{q^2
(q+p_1+p_2)^2}\int\frac{d^3 k}{(2\pi)^3}\frac{(2k+p_2)_\sigma
(2k+q+p_1+p_2)_\tau}{k^2(k+p_2)^2(k+q+p_1+p_2)^2}+.......\nonumber
\label{penul}
\end{eqnarray}
dots mean contributions to (\ref{master}), which doesn't play any
role in the computation of this Wilson coefficient.

After regularization of (\ref{penul}) the contribution to the
Wilson coefficient is equal to
\begin{equation}
\frac{7}{12}e^4\frac{1}{(q^2)^2\sqrt{q^2}}.
\end{equation}
The third type of diagrams gives:
\begin{eqnarray}
&&\frac{e^4}{6 p_1^2 p_2^2}\frac{1}{q^2(q+p_1+p_2)^2}
\int\frac{d^3 k}{(2\pi)^3}\nonumber\\
&& \left\{\frac{(2k+p_1)_\tau (2k-p_2)_\sigma
(2k+2p_1+q)\cdot(2k+q+p_1-p_2)}{k^2 (k+p_1)^2 (k+q+p_1)^2
(k-p_2)^2}
\right.\nonumber\\
&& \left.+\frac{(2k+p_1)_\tau (2k+2q+2p_1+p_2)_\sigma
(2k+2p_1+q)\cdot(2k+q+p_2+p_1)}{k^2 (k+p_1)^2 (k+q+p_1)^2
(k+q+p_1+p_2)^2}
\right.\nonumber\\
&& \left.+\frac{(2k+p_2)_\sigma (2k-p_1-2q)_\tau
(2k-q)\cdot(2k-q-p_1+p_2)}{k^2 (k+p_2)^2 (k-p_1-q)^2
(k-q)^2}\right\}\nonumber\\&&+\left\{p_1\leftrightarrow p_2\,\,
\bigwedge\,\, \sigma\leftrightarrow\tau\right\}.
\end{eqnarray}
Its contribution to Wilson coefficient is
\begin{equation}
-\frac{241}{192}e^4\int\frac{d^3 q}{(2\pi)^3}\frac{\exp(-\imath q
x)-1}{(q^2)^2\sqrt{q^2}}.
\end{equation}
Finally the Wilson coefficient in the momentum space for $<A^2>$
for three-dimension scalar quantum electrodynamics is
\begin{equation}
-\frac{75}{64}e^4\frac{1}{(q^2)^2\sqrt{q^2}}.
\end{equation}

\subsection{Note about the regularization of some divergent integrals}
The integration on $k$ in the previous section can be done by
using Feynman parameters. All integrand are integrable in the
infrared and in the ultraviolet but they have non-integrable
singularities for $k=q$. The renormalized values of these
integrals can be obtained by analytical continuation
\cite{vladimir}, which is applied in the following way. First one
introduces into the initial divergent integral a parameter in such
a way that for some values of this parameter that integral becomes
convergent. Certainly these finite values of modified integral
have no (direct) relation to that one of initial divergent
integral. However, if the modified integral is an analytic
function of the regularization parameter, that integral is defined
through this function in its analyticity domain. Further one
postulates that the renormalized (or physical) value of the
initial divergent integral is equal to the respective value of
this new analytic function. Let us consider the regularization of
the following integral as representative of the all others:
\begin{equation}
\int\frac{d^3 k}{(2\pi)^3}\frac{1}{k^2[(k-q)^2]^2} \label{primo}
\end{equation}
Due to the bad non-integrable singularity at $k=q$, I consider the
following integral depending on the complex parameter $h$
\begin{equation}
\int\frac{d^3 k}{(2\pi)^3}\frac{1}{k^2[(k-q)^2]^h} \label{primo}
\end{equation}
By introducing Feynman parameters:
\begin{equation}
\frac{1}{k^2((k-q)^2)^h}=
-\imath^{-2h}\frac{\Gamma(h+1)}{\Gamma(h)}\int_{0}^{1}\frac{x^{h-1}}
{(-k^2+2xkq-xq^2)^{h+1}}dx
\end{equation}
therefore (\ref{primo}) becomes
\begin{equation}
-\imath^{-2h}\frac{\Gamma(h+1)}{\Gamma(h)}\int_{0}^{1}dx\int\frac{d^3
k}{(2\pi)^3}\frac{x^{h-1}}{(-k^2+2xkq-xq^2)^{h+1}}
\end{equation}
The integration on $k$ gives:
\begin{equation}
-\imath^{-2h}\frac{1}{\Gamma(h)}
\int_{0}^{1}dx\frac{\Gamma\left(h-\frac{1}{2}\right)}{
(4\pi)^{3/2}}\frac{x^{h-1}}{(x^2q^2-x
q^2)^{h-\frac{1}{2}}}\label{secondo}
\end{equation}
and integrating on $x$
\begin{equation}
-\frac{\imath^{-4h+1}\pi}{(4\pi)^{3/2}
(q^2)^{h-1/2}\sin\left[\pi\left(h-\frac{1}{2}\right)\right]\Gamma(h)}
\frac{1}{\Gamma(h-2)}
\end{equation}
for $Re[h]<\frac{3}{2}$. By avoiding poles at seminteger values of
$h$, the value of this integral can be analytically continued in a
neighbourhood of $h=2$, and I conclude that the regularized value
of (\ref{primo}) is
\begin{equation}
\left[\int\frac{d^3
k}{(2\pi)^3}\frac{1}{k^2((k-q)^2)^2}\right]_{\rm reg.}=0.
\end{equation}

\section{Conclusions}
In this letter I have proved that photon propagator of
four-dimensional spinor electrodynamics and three-dimensional
scalar electrodynamics might receive contributions from the
condensate $<A^2>$ because its Wilson coefficient is not zero.
Previous works claimed the cancellation of the contribution of
this condensate in photon propagator due to a - not at all there-
gauge invariance of this Green function. This non-zero result is
therefore totally expected also from the non-Abelian analysis
provided by Lavelle et all.\cite{lavelle}. Here I have found that
the contribution of $<A^2>$ might affect also that part of the
photon propagator which enters directly in the construction of
gauge invariant physical amplitudes. In particular, by using the
plane wave method to calculate physical amplitudes, one realizes
that the condensate $<A^2>$ contributes to the amplitude of four
photons. It would be interesting to understand the possible role
played by $<A^2>$ in the phenomenology of the hard-scattering
light by light.

The contribution of $<A^2>$ to such a physical amplitude can be
considered a first experiment to check the gauge invariance of
this condensate. It will be the subject of next investigations to
compute directly this condensate in gauge theories in a non
trivial background, as for instance an instanton, in order to
check its gauge invariance beyond perturbation theory.

\section*{Acknowledgments}
This work is supported by a postdoctoral fellowship provided by
University of the Witwatersrand, Johannesburg. I acknowledge also
the South African National Research Foundation (NRF) for partial
financial support.


\begin{thebibliography}{0}
\bibitem{Jackiw}R.Jackiw, S.Templeton, {\it Phys. Rev.} {\bf D23}
(1981) 2291.
\bibitem{Pisarsky}T.Appelquist, R.Pisarski, {\it Phys.Rev.} {\bf
D23}(1981) 2305.
\bibitem{slavnov}A.A.Slavnov, {\it Phys.Lett.} {\bf B608} (2005) 171.
\bibitem{wilson}K.G.Wilson,{
\it Phys. Rev.} {\bf 179} (1969) 1499.
\bibitem{vari}M.A.Shifman, A.I.Vanshtein and V.I.Zakharov,
{\it Nucl. Phys.}{\bf B147} (1979) 385, 488;\\
L.J.Reinders, H. Rubinstein and S. Tazaki, {\it Phys.Rep.} {\bf
127} (
1985) 1;\\
S. Narison, {\it QCD, spectral sum rules}, World
Scientific, (1989);\\
J.C. Collins, {\it Renormalization}, Cambridge University Press,
(1984).
\bibitem{scadron}V.Elias, T.G.Steele, M.D.Scadron, {\it Phys.Rev.}
{\bf D38} (1988) 1584;\\
E.Bagan, T.G.Steele, {\it Phys. Lett.} {\bf B219} (1989) 497.
\bibitem{QED}A.I. Akhiezer, V.B. Berestetskii,
{\it Quantum Electrodynamics}, Interscience Publishers (1965).
\bibitem{vladimir}see for instance:
G.Lambiase, V.V. Nesterenko, M.Bordag, {\it Journ. Math. Phys.}
{\bf 40} (1999) 6254.
\bibitem{lavelle}M.Lavelle, M.Schaden, {\it Phys. Lett.} {\bf
B208} (1988) 207.
\end{thebibliography}
\end{document}